\documentclass[]{emulateapj}

\usepackage{epsfig,amsmath,amssymb,amsfonts,mathrsfs,latexsym,graphicx,lscape}
\usepackage{hyperref}

\begin{document}

\title{Discovery of high-frequency quasi-periodic oscillations in the
  black-hole candidate IGR~J17091--3624}

\author{D. Altamirano\altaffilmark{1} \& T. Belloni\altaffilmark{2}}

\altaffiltext{1}{Email: d.altamirano@uva.nl ; Astronomical Institute,
  ``Anton Pannekoek'', University of Amsterdam, Science Park 904,
  1098XH, Amsterdam, The Netherlands}

\altaffiltext{2}{INAF-Osservatorio Astronomico di Brera, Via
  E. Bianchi 46, I-23807 Merate (LC), Italy}

\begin{abstract}

We report the discovery of $8.5\sigma$ high-frequency quasi-periodic
oscillations (HFQPOs) at 66~Hz in the RXTE data of the black hole
candidate IGR~J17091--3624, a system whose X-ray properties are very
similar to those of microquasar GRS~1915+105. The centroid frequency
of the strongest peak is $\sim$66 Hz, its quality factor above 5 and
its rms is between 4 and 10\%.
 We found a possible additional peak at 164 Hz when selecting a subset
 of data; however, at 4.5$\sigma$ level we consider this detection
 marginal.
These QPOs have hard spectrum and are stronger in observations
performed between September and October 2011, during which
IGR~J17091--3624 displayed for the first time light curves which
resemble those of the $\gamma$ variability class in GRS~1915+105.
We find that the 66~Hz QPO is also present in previous observations
(4.5$\sigma$), but only when averaging $\sim$235~ksec of relatively
high count rate data.
The fact that the HFQPOs frequency in IGR~J17091--3624 matches
surprisingly well that seen in GRS~1915+105 raises questions on the
mass scaling of QPOs frequency in these two systems. 
We discuss some possible interpretations, however, they all strongly
depend on the distance and mass of IGR~J17091-3624, both completely
unconstrained today.

\end{abstract}

\section{Introduction}\label{sec:intro}

One of the strongest motivations for studying Low-mass X-ray binaries
(LMXBs) has been the aim to use these systems as probes of fundamental
physics. 
It has been argued that the study of high frequency (40--450~Hz)
quasi-periodic oscillations (HFQPOs) found in LMXBs containing a
stellar mass black hole (BH) is one of the most promising tools to do
so.
The HFQPOs are the fastest phenomenon observed in these systems and
their high frequency suggests that they are produced in the innermost
region of the accretion flow near the BH event horizon.
Once the processes involved in their formation are firmly
understood, they will give a wealth of information on BH mass
and spin as well as the structure of strongly curved spacetime
\citep[e.g.,][]{Nowak97, Wagoner99, Stella99a,  Wagoner01, Abramowicz01, Kato01,
  Rezzolla03}.

\begin{figure*} 
\centering
\resizebox{2\columnwidth}{!}{\rotatebox{-90}{\includegraphics[clip]{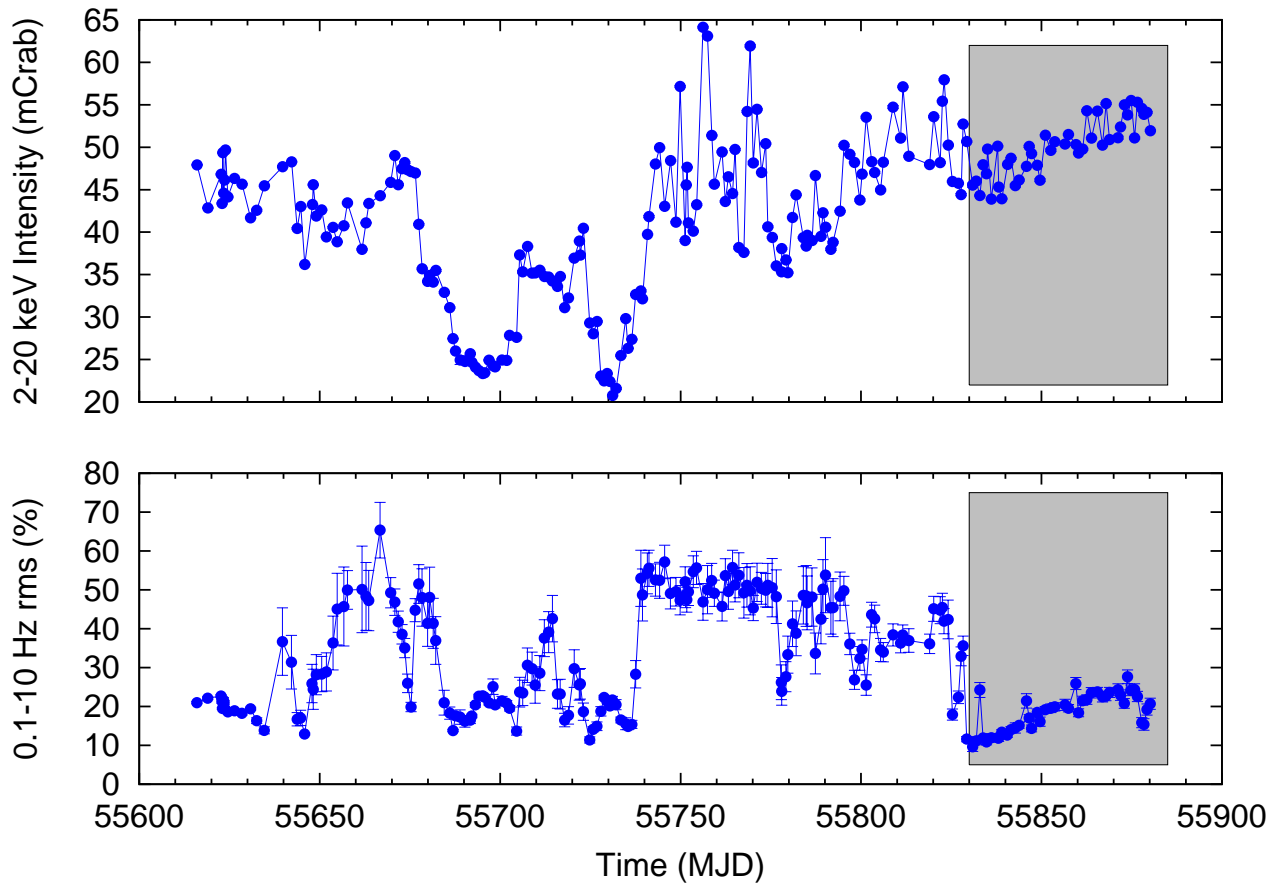}}}
\caption{\textit{Upper panel:} Crab-normalized 2-20 keV intensity of
  the full outburst of IGR~J17091--3624 sampled with
  RXTE. \textit{Lower panel:} 0.01-10 Hz fractional rms amplitude
  (2-60 keV) range. Points are per observation.  The grey area marks
  the period of fast $\gamma$ class of variability
  \citep[e.g.,][]{Belloni00}. The average power spectra of all
  observations in this region is shown in Figure~\ref{fig:pds}.  }
\label{fig:lc}
\end{figure*}

The frequency of the HFQPOs usually occur at specific values,
different in each source. In a few cases, pairs of QPOs have been
detected. The highest frequency appears to scale inversely
proportional to the to BH mass \citep[e.g.,
][]{McClintock06,Belloni06a}.
HFQPOs are weak, transient, and energy dependent, and are mostly
detected only at high count rates.  In some cases they are found just
above detection levels, so there is considerable uncertainty about
their exact properties. However, typically the quality factor Q ranges
between less than 2 and a few tens, and amplitudes are between 0.5 and
5\% rms (depending on the source and the energy range used).

Twin QPOs at a stable 300 Hz and 450 Hz \citep{Strohmayer01a} have
been observed in the BH GRO J1655--40; at 240 and 160 Hz in
H1743--322 \citep{Homan05b} and at 188 and 268 Hz in XTE J1550--56
\citep[note however, that in this case it is not clear how stable the
  frequency of these QPOs is, see][]{Remillard99,Homan01,Miller01}.
Single QPOs or peaked noise (i.e. when $Q\lesssim$2) features at a
rather stable frequency have been detected in XTE~J1650--500
\citep[$\sim$250~Hz, see, e.g,][]{Homan03}, 4U~1630--47
\citep[variable frequency, see ][]{Klein04} and XTE J1859+226
\citep[$\sim$180~Hz, see, e.g., ][]{Cui00a}.

The remaining BH which shows HFQPOs is the peculiar system
GRS~1915+105, which is a 33.5 days orbital period binary system
harboring a $14.0\pm4.4 \ M_{\odot}$ BH \citep{Greiner01,
  Harlaftis04}. At a distance of $\sim12.5$ kpc \citep{Mirabel94}, it
is very often at Eddington or super-Eddington luminosity
\citep[e.g.,][]{Done04}.
Until recently, GRS~1915+105 has been unique in that its X-ray light
curves exhibit more than a dozen different patterns of variability
usually called ``classes'' (which are referred to with Greek letters),
most of which are high amplitude and highly-structured
\citep[e.g.,][]{Belloni00}.
Most of this structured variability occurs on timescales of seconds or
longer, and it is thought to be due to limit cycles of accretion and
ejection in an unstable disk \citep[e.g.,][]{Belloni97, Mirabel98,
  Tagger04, Neilsen11}.

GRS~1915+105 has HFQPOs at the rather low frequencies of $\sim$40 Hz
\citep{Strohmayer01b} and $\sim$67~Hz \citep{Morgan97} compared
with the rest of the BH systems; however, as in the case of
GRO J1655--40 and H1743--322, the QPO frequencies remain almost
constant as the X-ray flux changes (any change in the frequency of the
$\sim$40 Hz oscillation during the sequence of observations was
$\lesssim$1\%; the frequency of the 67 Hz oscillation varies by only a
few percent as the X-ray flux varies by a factor of several),
suggesting a common origin between systems.
\citet{Belloni01} reported on the detection of QPOs at $\sim$27~Hz;
given the weakness of the signal \citet{Belloni01} could not determine
whether the $\sim$27~Hz was related to the $\sim$67~Hz QPO (at the
time of \citealt{Belloni01}, the $\sim$40 Hz QPO was not yet
reported); however, they were able to conclude that the
presence/absence of HFQPOs is intimately linked to the slower
oscillations and variations that happen on timescales longer than
seconds.  An additional peak at $\sim$170 Hz was also found after
applying a very specific class-count rate-hardness selection
\citep{Belloni06a}; however, this barely significant peak is very
broad, with a Q value or $\sim$2.

\begin{figure} 
\centering
\resizebox{1\columnwidth}{!}{\rotatebox{0}{\includegraphics[clip]{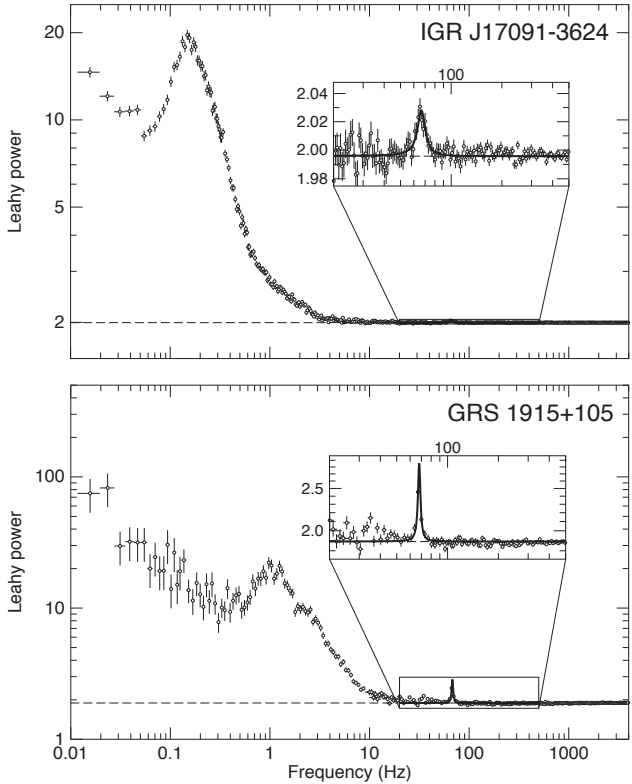}}}
\caption{\textit{Upper panel}: Averaged power spectra of all
  observations of IGR~J17091--3624 marked by the grey area in
  Fig.~\ref{fig:lc} (for a total of $\sim$74~ksec of data).
  \textit{Lower panel}: Average power spectra of a single observation
  (ObsID: 80701-01-28-01) of GRS~1915+105. Both plots show the typical
  $\gamma$ variability class power spectra, with a strong broad bump
  at low frequencies.  Insets show a zoom to the $\sim65-67$~Hz QPO. }
\label{fig:pds}
\end{figure}

\begin{figure*} 
\centering
\resizebox{2\columnwidth}{!}{\rotatebox{0}{\includegraphics[clip]{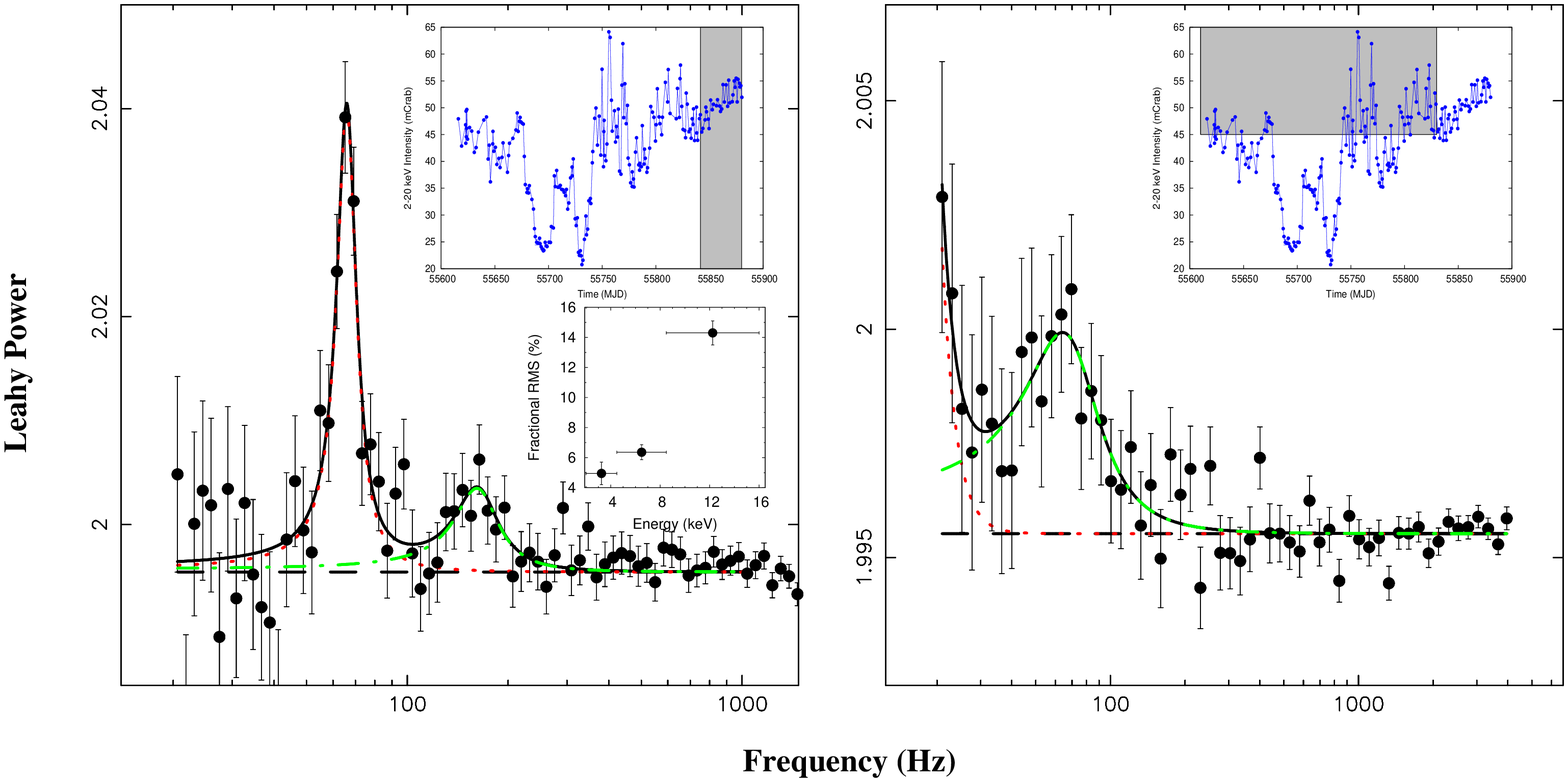}}}
\caption{\textit{Left panel:} Average power spectrum of all
  observations after MJD 55841 (for a total of 48~ksec). The
  $66\pm0.5$~Hz QPO is 9$\sigma$ significant, while the
  $\sim164\pm10$~Hz is 4.5$\sigma$. Lower inset shows the rms
  amplitude vs. energy for the 66~Hz QPO. \textit{Right panel:}
  Average power spectrum of all observations performed before MJD
  55830, when the average intensity of IGR~J17091--324 was higher than
  46 mCrab (corresponding to $\sim$100 cts s$^{-1}$ PCU$^{-1}$; we
  used a total of $\sim$235~ksec). Note the difference the \textit{y}
  axis scales. The broad (Q=1) QPO at $70\pm6$ is 4.5$\sigma$
  significant. In both panels we used data in the 2-25 keV
  range. Insets on the top: light curves marking with grey area the
  data used to produce each power spectrum. }
\label{fig:pdsselect}
\end{figure*}

If the frequency of the HFQPOs depends on the mass and/or spin of the
BH, one would expect (as observations appear to suggest), that
different systems with different parameters should show QPOs at
different frequencies.
The recent discovery of a BH (IGR~J17091--3624) which shows
very similar high amplitude and highly-structured variability to that
seen in GRS~1915+105 \citep[see][ and references
  therein]{Altamirano11d} opened a new window of opportunities to
understand the physical mechanism that produces the highly structured
X-ray variability and it relation with much faster (sub-second)
variability.

IGR~J17091--3624 has been seen active in 1994, 1996, 2001, 2003, 2007
and 2011 \citep{Revnivtsev03a, Intzand03, Capitanio06, Kuulkers03b,
  Capitanio09, Krimm11}.
Only in the recent 2011 outburst IGR~J17091--3624 started to show the
high amplitude and highly-structured variability \citep{Altamirano11a,
  Altamirano11b, Altamirano11c, Altamirano11d, Pahari11} typical of
GRS~1915+105.
In the context of the variability classes defined by \citet{Belloni00}
for GRS~1915+105, \citet{Altamirano11d} found that IGR~J17091--3624
shows the $\nu$, $\rho$, $\alpha$, $\lambda$, $\beta$ and $\mu$
classes as well as quiet periods which resemble the $\chi$ class, all
occurring at 2-60 keV count rate levels which can be 10-50 times lower
than observed in GRS~1915+105.
\citet{Altamirano11d} also found that the difference in flux combined
with the circumstance that currently neither the distance to
IGR~J17091-3624 nor the mass of its compact object are known lead to
the conclusion that either all models requiring near Eddington
luminosities for GRS~1915+105-like variability fail, or
IGR~J17091--3624 lies at a distance well in excess of 20 kpc or, it
harbors one of the least massive BHs known ($< 3 M_\odot$).

The similarities and differences between IGR~J17091--3624 and
GRS~1915+105 raises the question of at what frequency should we
observe the HFQPOs in IGR~J17091--3624. Assuming 1/M scaling, and the
possibility that IGR~J17091--3624 has a mass which is a factor of
$\sim$5 smaller than GRS~1915+105, then the HFQPOs in IGR~J17091--3624
should be at a frequency significantly higher than the $\sim$40~Hz and
$\sim$67~Hz seen in GRS~1915+105. Finding similar HFQPO frequency
between sources would imply that the frequency does not depend on the
mass or that IGR~J17091--3624 lies at a distance well in excess of 20
kpc.

Triggered by this question, we searched all the \textit{Rossi X-ray
  Timing Explorer} (RXTE) available data of IGR~J17091--3624 for the
presence of QPOs at frequencies higher than the $\sim$7-10~Hz reported
by \citet{Altamirano11d}.
In this Letter we report the discovery of a highly significant QPO at
$\sim$65~Hz, and a marginally detection of a QPO at $\sim$165~Hz and
briefly discuss the implication of our detections.

\section{Observations and data analysis}\label{sec:observations}

IGR~J17091--3624 was observed with the Proportional Counter Array
\citep[PCA; ][]{Jahoda06} on-board RXTE almost daily since the
outburst began in February 2011 until October 15th, 2011, when RXTE
could not point to IGR~J17091--3624 any more due to visibility
constraints.  A total of 228 observations, covering $\sim280$ days,
sample the 2011 part of the current outburst (At the moment of
submission it is not known whether the source is still active or not).
In this work we do not use data taken between MJD~55600 and 55614 (14
observations), given that these observations were affected by a bright
source in the field of view \citep[e.g.,][]{Altamirano11d}.

We use the 16-s time-resolution Standard 2 mode data to calculate the
Crab-normalized average per observation 2--20 keV intensity
\citep[see, e.g., ][and references therein for details on the
  method]{Altamirano08}. 
Power spectra and light curves were produced from the PCA using
standard techniques \citep[e.g.,][]{Belloni00,Altamirano08}. For the
Fourier timing analysis we used data in the 2-25 keV range (absolute
channels 5--60) from the Good Xenon mode.

To fit the high-frequency ($>10$~Hz) part of power spectra, we used
one or two Lorentzian functions plus a power-law when needed.  We give
the frequency of the Lorentzians in terms of characteristic frequency
$\nu_{max} = \sqrt{\nu^2_0 +(FWHM/2)^2} = \nu_0 \sqrt{1 + 1/4Q^2}$
 \citep{Belloni02}.  For the quality factor $Q$ we use the
standard definition $Q = \nu_0/FWHM$. \textit{FWHM} is the full width
at half maximum and $\nu_0$ the centroid frequency of the Lorentzian.
The strength of the QPOs are given in terms of fractional rms
amplitude.

\section{Results}\label{sec:results}

\subsection{Outburst evolution}

In the upper and lower panel of Fig.~\ref{fig:lc} we show the
average 2--20~keV intensity for each observation and the 0.01-10~Hz
rms amplitude, respectively.
During the 265 days period, IGR~J17091--3624 has been at an average
intensity between 20 and 65 mCrab. (Note that during single flares or
heartbeats the source was much brighter).
The rms amplitude also varied significantly, where
IGR~J17091--3624 underwent periods of very low ($\sim$10\%) and very
high ($>$60\%) amplitude.
Figure~\ref{fig:lc} shows that there is no clear correlation between
intensity and rms amplitude. A similar study of
GRS~1915+105 shows the same (this will be reported elsewhere).

\subsection{High frequency QPOs}\label{sec:kHz}

At first we searched each observation power spectrum for indications
of QPOs at higher frequency than 10~Hz and at a significance larger
than 3.5$\sigma$.
We found a clear detection (4.7$\sigma$, single trial) of a QPO at
$\nu=64.8\pm1.7$~Hz, $Q=4.8\pm1.7$ and rms amplitude of
$10\pm1$\% (2-25 keV, ObsID: 96420-01-36-03, MJD 55865.5).
We also found that in the period MJD~55830-55880 many power spectra
revealed power excesses in the 60-70~Hz range. To further investigate
this, we added all data in this period , which corresponds to the
low-variability period marked with a grey area in Fig.~\ref{fig:lc}.
In the upper panel of Fig.~\ref{fig:pds} we show that the resulting
power spectrum is clearly dominated by a strong feature at
$\sim$0.18~Hz, but also shows an 8.5$\sigma$ QPO at $66.5\pm0.5$~Hz
($Q=7.8\pm1.5$; $4.9\pm0.2$\% rms amplitude).

Different sub-selections on time and intensity of these data lead to
significant changes in Q, amplitude and significance, but not in
frequency. This suggests that that the 66~Hz QPO is not always
present, similar to what occurs in GRS~1915+105.
In several of our sub-selections we saw an excess of power in the
150-200 Hz range.
In the left panel of Fig.~\ref{fig:pdsselect} we show the result of
averaging all observations after MJD~J55841. The QPO at $\sim66$~Hz
($9\sigma$ significant, $Q=6\pm1$, $6.4\pm0.4$\% rms amplitude) is
accompanied with a a QPO at $164\pm10$~Hz ($\sim$4.5$\sigma$,
$Q=2.6\pm1.3$, $6.6\pm0.8$\% rms amplitude);
In the inset we show the rms amplitude vs. energy of the 66~Hz QPO;
its amplitude increases with energy from $\sim$5\% at 3 keV, to
$\sim$14\% at 13 keV.
The 67~Hz QPO in GRS~1915+105 also has a hard spectrum, although less
extreme than this one \cite{Morgan97}.  The low S/N did not allow a
similar analysis for the 164~Hz QPO.

We also explored whether the 65~Hz QPO is also present before
MJD~55830 which is not detected due to the low-statistics. Averaging
all data in this period led to a broad feature at $\sim$60-80 Hz whose
parameters were unconstrained.
When we selected only high count rates ($>110$ cts/sec/PCU), we
detected a 4.5$\sigma$ broad ($Q=1.2\pm0.5$) feature at $70\pm6$~Hz
($4.4\pm0.4$\% rms amplitude).
We show our best fit n the right panel of Fig.~\ref{fig:pdsselect}.

We did not detect any QPO at 40~Hz nor 27~Hz as seen in
GRS~1915+105. This is not surprising, as those QPOs are can be weaker
and complex data selection might be needed to find them
\citep[e.g.,][]{Strohmayer01b,Belloni01}

\begin{figure} 
\centering
\resizebox{1\columnwidth}{!}{\rotatebox{0}{\includegraphics[clip]{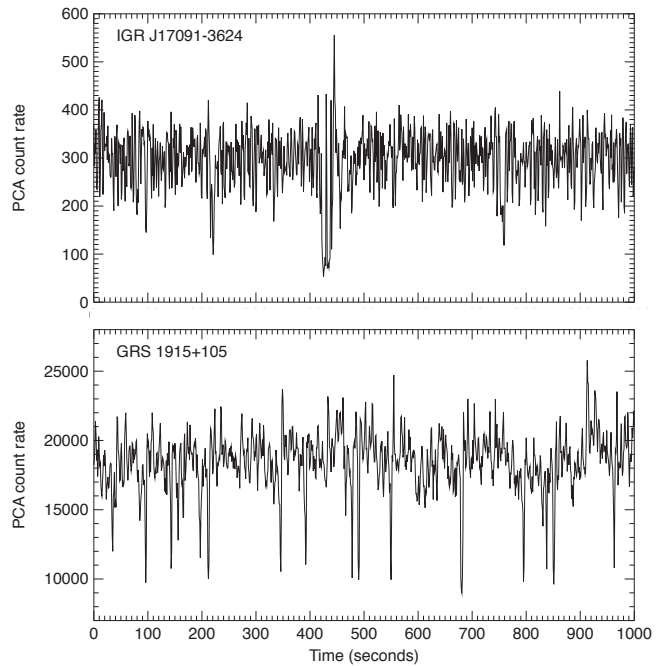}}}
\caption{Upper and lower panel show 1 sec bin light curves of
  IGR~J17091--3624 (subset of ObsID 96420-01-37-04) and GRS~1915+105
  \citep[subset of observation 20402-01-39-00, see also][]{Belloni00},
  respectively, during the $\gamma$ variability class.}
\label{fig:lccomp}
\end{figure}

\subsection{The low-variability period MJD~55830-55880}

The period MJD~55830-55880 (where the 66~Hz QPO is strongest, see
Section~\ref{sec:kHz}) corresponds to an interval of relative low
(10-30\%) rms amplitude.
The 1-sec light curves during this period are relatively flat, apart
from quasi-periodic oscillations with a typical time scale of 10 s
and/or the presence of sharp ‘dips’ with a typical duration of a few
seconds.
In the upper panel of Fig.~\ref{fig:lccomp} we show a representative
lightcuve during this period.

When compared with the GRS~1915+105 variability classes defined by
\citet{Belloni00}, the IGR~J17091--3624 light curves during
MJD~55830-55880 are reminiscent of the $\gamma$ class. As a
comparison, in the lower panel of Fig.~\ref{fig:lccomp} we plot a
representative $\gamma$ class data-segment for GRS~1915+105 \citep[for
  the full light curve, see Figs. 2f and 15b,d in][]{Belloni00}.
A comparison between power spectra between the IGR~J17091--3624 and
GRS~1915+105 is shown in Fig.~\ref{fig:pds}.

\section{Discussion}

We discovered the presence of a HFQPO in the RXTE data of the
IGR~J17091--3624, a system whose X-ray properties are very similar to
those of GRS 1915+105. The centroid frequency of the strongest peak is
$\sim$66 Hz, its quality factor above 5 and its rms is between 4 and
10\%. We found a possible additional peak at 164 Hz when selecting a
subset of data; however, we consider this detection marginal.

The HFQPOs are detected during a period where IGR~J17091--3624 is in
the $\gamma$ variability class \citep{Belloni00}.
\citealt{Altamirano11d} reported observations until MJD~55800 where so
far IGR~J17091--3624 was seen in at least the $\nu$, $\rho$, $\alpha$,
$\lambda$, $\beta$, $\mu$ and $\chi$ classes.
The fact that now we also identified the $\gamma$ class probably means
that IGR~J17091--3624 is still evolving between classes.

%%%%%%%%%%%%%%%%%%%%%%%%%%%%%%%%%%%%%%%%%%%%%%%%%%%%%%%%%%%%%%%%%%%
%%%%%%%%%%%%%%%%%%%%%%%%%%%%%%%%%%%%%%%%%%%%%%%%%%%%%%%%%%%%%%%%%%%
%%%%%%%%%%%%%%%%%%%%%%%%%%%%%%%%%%%%%%%%%%%%%%%%%%%%%%%%%%%%%%%%%%%
%%%%%%%%%%%%%%%%%%%%%%%%%%%%%%%%%%%%%%%%%%%%%%%%%%%%%%%%%%%%%%%%%%%
%%%%%%%%%%%%%%%%%%%%%%%%%%%%%%%%%%%%%%%%%%%%%%%%%%%%%%%%%%%%%%%%%%%
%
% Discussion the different possibilites
%

The frequency of the 66 Hz QPO in IGR~J17091--3624 is surprisingly
similar to that of the main HFQPO in GRS 1915+105 \citep[65-69 Hz,
  see][]{Morgan97}.
In addition, the frequency of the 4.5$\sigma$ significant broad
feature at $164\pm10$~Hz in IGR~J17091--3624 also matches that of a
similar broad feature found at 170~Hz in GRS~1915+105
\citep{Belloni06}.
The circumstance that currently neither the distance to
IGR~J17091-3624 nor the mass of its compact object are known led
\citet{Altamirano11d} to the conclusion that either all models
requiring near Eddington luminosities for GRS~1915+105-like
variability fail, or IGR~J17091--3624 lies at a distance well in
excess of 20 kpc or, it harbors one of the least massive BHs
known ($< 3 M_\odot$).
The surprisingly match between the $\sim$70~Hz and $\sim$170~Hz
frequency therefore raises more questions than answers.

The frequencies of dynamical motions near compact objects are thought
to scale with mass \citep[e.g.,][]{Vanderklis06}. 
The slower highly-structured low-frequency variability seen in
GRS~1915+105 as compared with IGR~J17091-3624 supported the idea that
the later was less massive \citep{Altamirano11d}. However, the fact
that the HFQPOs occur at the same frequency could be at odds, meaning
that either the frequency of the HFQPOs or the highest-frequency of
the `heartbeats' in these two sources scale with mass, or that none of
them do, or that they both do thanks to a difference in BH
spin.

%%%%%%%%%%%%%%%%%%%%%%%%%%%%%%%%%%%%%%%%%%%%%%%%%%%%%%%%%%%%%%%%%%%
%%%%%%%%%%%%%%%%%%%%%%%%%%%%%%%%%%%%%%%%%%%%%%%%%%%%%%%%%%%%%%%%%%%
%
% If the frequency scales with mass ...
%

If the 67~Hz HFQPOs frequency scales with mass (assuming same mass,
same spin), then either IGR~J17091--3624 is further than 20~kpc or it
is close, in which case all models requiring near Eddington
luminosities for GRS~1915+105-like variability fail
\citep{Altamirano11d}.
Under the assumption that the centroid frequencies represent that of a
Keplerian circular motion at the inner-most stable orbit (ISCO), then
at first approximation \citep[e.g.,][]{Kluzniak90} no solution is
possible for the 67~Hz QPOs independently of the spin, unless either
the HFQPOs are not produced at the ISCO or the mass of both BHs is
significantly higher than the $\sim$14$M_\odot$ estimated for
GRS~1915+105.
The 160-170~Hz broad QPOs found in both in IGR~J17091--3624 (this
Letter) or GRS~1915+105 would be consistent with a 14-20$M_\odot$, but
only if the spin parameter ($a/M_{BH}$) in these systems is
significantly lower than 1 \citep[i.e., at odds with the reported
  $a/M_{BH} \simeq 1$ for GRS~1915+105,
  e.g.,][]{McClintock06a,Blum09}.

In the framework of the Relativistic Precession Model
\citep[RPM,][]{Stella99a}, and relaxing the assumption on equal spin,
we can identify the higher frequency (160-170 Hz) as representing a
Keplerian motion at a certain radius and, the lower frequency (66~Hz)
as the relativistic periastron precession at the same radius (notice
that this radius must be larger than the ISCO, where the two
frequencies are the same).
To test this identification, we searched for the combinations between
spin parameter and BH mass which allow a solution. We find
solutions for masses ranging between 7~M$_\odot$ (for $a/M_{BH}=0$)
and 19~M$_\odot$ (for $a/M_{BH}=1$), the latter consistent with the
$14.0\pm4.4 \ M_{\odot}$ and $a/M_{BH}\simeq 1$ reported for
GRS~1915+105 \citep{Greiner01, Harlaftis04,McClintock06a,Blum09}.
This identification implies that IGR~J17091--3624 could be
considerably less massive than GRS~1915+105 (and coincidentally have a
much lower spin than GRS~1915+105), explaining in this way why the
highest frequency of the 'heartbeats' is significantly higher in
IGR~J17091--3624 than in GRS~1915+105 \citep{Altamirano11d}.
We note that in the framework of the Relativistic Precession Model the
41~Hz frequency would be unexplained. Solutions to the 41,67 Hz pair
for GRS 1915+105 are possible, but only for masses larger than
25M$_\odot$ and considering no spin (i.e $a/M_{BH}=0$).
%

%%%%%%%%%%%%%%%%%%%%%%%%%%%%%%%%%%%%%%%%%%%%%%%%%%%%%%%%%%%%%%%%%%%
%%%%%%%%%%%%%%%%%%%%%%%%%%%%%%%%%%%%%%%%%%%%%%%%%%%%%%%%%%%%%%%%%%%
%
% If the frequency DOES NOT scale with mass ...
%

If the HFQPOs frequency \textit{does not} scale with mass, then it is
intriguing why their HFQPOs frequencies are the same in these two
sources. One possibility is that the 67~Hz HFQPOs are different from
the HFQPOs seen in other sources \citep[e.g., ][]{Vanderklis06}, and
that their frequency is instead related with the physical process that
produces the highly-structured low-frequency variability (only seen)
in these two sources.
If true, this could have major implications on works that use the
$\sim67$~Hz HFQPO in GRS~1915+105 as a standard to estimate the mass
of galactic and/or super massive BHs
\citep[e.g. ][]{Middleton10}.

%%%%%%%%%%%%%%%%%%%%%%%%%%%%%%%%%%%%%%%%%%%%%%%%%%%%%%%%%%%%%%%%%%%
%%%%%%%%%%%%%%%%%%%%%%%%%%%%%%%%%%%%%%%%%%%%%%%%%%%%%%%%%%%%%%%%%%%
%%%%%%%%%%%%%%%%%%%%%%%%%%%%%%%%%%%%%%%%%%%%%%%%%%%%%%%%%%%%%%%%%%%
%%%%%%%%%%%%%%%%%%%%%%%%%%%%%%%%%%%%%%%%%%%%%%%%%%%%%%%%%%%%%%%%%%%
%%%%%%%%%%%%%%%%%%%%%%%%%%%%%%%%%%%%%%%%%%%%%%%%%%%%%%%%%%%%%%%%%%%
%
% ending the paper
%

Our results add another HFQPO to the small number already known. Its
properties are not different from the others, in particular its
elusiveness. There are thousands of observations of more than 20
black-hole transients in the RXTE archive, but only very few
significant peaks have been found. As RXTE will cease operations in
2012 January, this sample is not going to be increased until the
Indian astronomical mission ASTROSAT is launched, while a major step
forward will only come with the next generation of X-ray instruments
\citep[e.g. LOFT, see][]{Feroci11}.

\textbf{Acknowledgments:} TB acknowledge support from grant INAF-ASI
I/009/10/0. The research leading to these results has received funding
from the European Community’s Seventh Framework Programme
(FP7/2007-2013) under grant agreement number ITN 215212 “Black Hole
Universe”.

\end{document}